# BUCKET SHAKING STOPS BUNCH DANCING IN TEVATRON

A. Burov and C. Y. Tan, FNAL[*]

*Abstract*

Bunches in Tevatron are known to be longitudinally unstable: their collective oscillations, also called "dancing bunches", persist without any signs of decay [1]. Typically, a damper is used to stop these oscillations, but recently, it was theoretically predicted that the oscillations can be stabilized by means of small bucket shaking [2,3]. Dedicated measurements in Tevatron have shown that this method does stop the dancing.

## INTRODUCTION

The spectrum of the Boltzmann-Jeans-Vlasov (BJV) [4] equation consists of continuous and, possibly, discrete parts [5,6]. The emergent coherent mode without any sign of decay is the discrete van Kampen mode. Due to this loss of Landau damping (LLD), even a tiny coupled-bunch wake is sufficient to drive an instability. Thresholds of LLD are rather low and are strongly dependent on the bunch distribution function $F(I)$ [3]. In particular, for a full bucket of a single-harmonic RF and inductive impedance, the threshold incoherent low-amplitude synchrotron tune shift can be as low as 10% for the Hofmann-Pedersen distribution, and just ~1% for a model coalesced bunch. In terms of the bunch population, the two thresholds differ almost by two orders of magnitude. It turns out, that LLD onset is highly sensitive to the steepness of the distribution function at low amplitudes: the flatter the distribution, the more stable it is. This prediction appears to be generally correct when the bare RF synchrotron frequency monotonically decreases with the amplitude and effectively repulsive wake field, diminishing the incoherent synchrotron frequencies. For example, in the case of a sinusoidal RF, any combination of inductance, wall resistivity, or high order modes above transition, or space charge below transition shifts the incoherent spectrum down to lower frequency. It is important that the coherent frequency is not moved as much: for example, in the case of linear RF focusing, the coherent frequency is not moved at all; it stays equal to the bare RF synchrotron frequency at any beam current [3,7]. As a result, the coherent mode emerges above the incoherent spectrum and thus, it is mostly associated with the low-amplitude particles. This mode looks like dipole motion of some central portion of the bunch – exactly how it was observed at Tevatron [1]. From this point of view, the LLD threshold can be estimated from the threshold for a rigid-bunch mode applied to a distribution with an effective bunch length $l$. According to the LLD power law found in Ref. [8], the threshold bunch population $N_{th}$ is a strong function of $l$: for the inductive wake above transition $N_{th} \propto l^5$. This scaling follows from the idea that LLD happens when the incoherent synchrotron tune shift $\Delta\Omega \propto NZ_\parallel(l^{-1})/l^2$ exceeds the incoherent tune spread $\delta\Omega \propto l^2$. Thus, the smaller the excited central portion of the distribution, the lower the threshold for instability. Since the coherent mode is associated with the low-amplitude particles, $l$ should be sensitive to the steepness of the bunch distribution function at low amplitudes, i.e. the steeper the distribution, the smaller the effective length $l$, and thus, the lower is the LLD threshold. These qualitative considerations are fully supported by the quantitative analysis of Ref. [3]. Thus, flattening of the bunch distribution at low amplitudes should make the bunch more stable.

## BUCKET SHAKING

To flatten the bunch distribution at small amplitudes, resonant shaking of the RF phase can be used [2]. Indeed, if the RF phase is modulated at the synchrotron frequency of the low-amplitude particles, these particles are captured by a first-order resonance separatrix, which has a width that is regulated by the shaking amplitude. After a sufficient number of synchrotron periods, the distribution function inside that separatrix can be expected to get flatter. To prevent excitation of the tail particles, the shaking should be switched on and off gradually enough. Numerical simulation of that distribution modification was run with the following map:

$$z_{n+1} = z_n + \Delta t \cdot p_n;$$
$$p_{n+1} = p_n - \Delta t \cdot \sin(z_{n+1} - \varphi_0 \sin(t_n)); \quad (1)$$
$$t_{n+1} = t_n + \Delta t.$$

Here $z$ and $p$ are the coordinate and momentum in proper units, $\Delta t$ is the time step, $\varphi_0$ is the shaking amplitude. Simulated particle distribution functions (PDF) in action and phase domains, before and after shaking, are shown in Figs. (1,2) for $\varphi_0 = 0.05$ and simulation time $T_{sim} = 2\pi \times 90$ rad. It can be seen that the action distribution has been successfully flattened; there is a little divot which develops after shaking. The final phase distribution has a dipole modulation (2 $\pi$). This modulation can be expected to smear after several consecutive shakings.

---

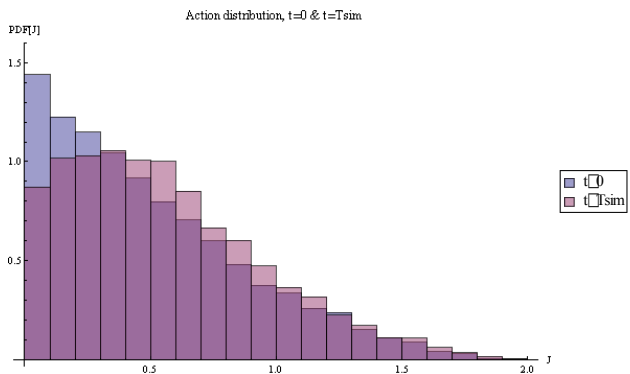

Figure 1: Initial (blue) and final (pink) distributions over action. Overlapping area is violet.

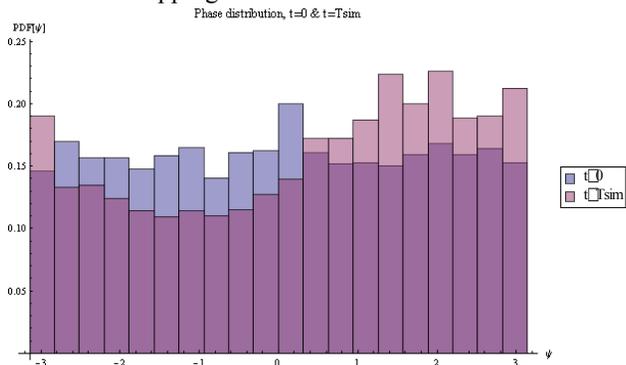

Figure 2: Same for the synchrotron phase distribution.

## EXPERIMENT

The hardware setup is shown in Figure 3. A signal generator and a phase shifter are used to phase modulate the Tevatron RF. The modulation frequency has been set to 34 Hz which is the synchrotron frequency at flattop and the amplitude of the sine wave has been set to 3°. In every shake, the amplitude of the modulation is ramped in the manner shown in Figure 4. This ramp has been chosen because it does not have any abrupt RF changes. Previous experiments have shown that any sudden turn on causes beam loss.

In the experiment, two bunches of protons are accelerated to flattop (980 GeV) and shaken five times there. The work has been done at flattop because the bucket area is much larger than the longitudinal beam emittance and thus allows the beam to freely change its shape without being constrained by the bucket edges. After every shake the bunch shapes are recorded with a 2 GHz bandwidth scope.

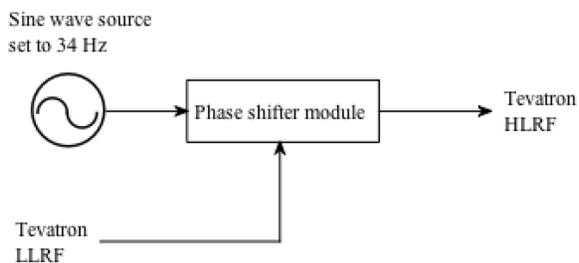

Figure 3: The hardware setup. A signal generator and a phase shifter is used to phase modulate the LLRF before it is sent to the HLRF.

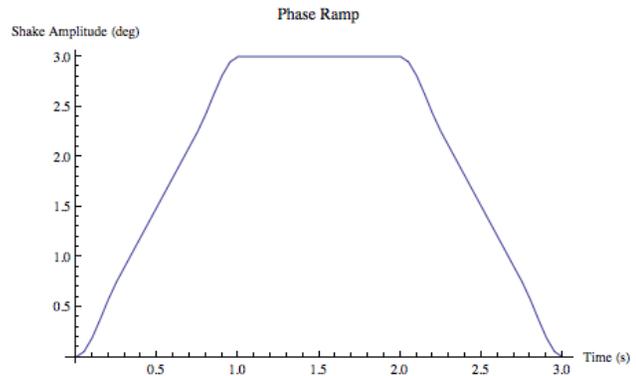

Figure 4: This is the ramp used for phase modulation.

*Results*

Figure 5 shows a data logger plot for the duration of the experiment. In the beginning, the bunch centroid (red trace) is moving a lot but at the end of five shakes, the centroid of the beam is clearly moving much less than before. There is also some bunch length growth (yellow trace) but there is no beam loss (green trace).

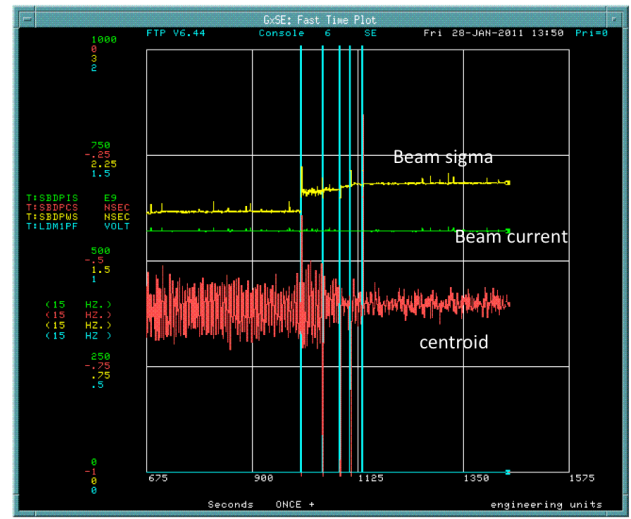

Figure 5: The beam is shaken five times (cyan spikes). After five shakes the beam centroid is not moving as much. There is some growth in bunch length but there is no beam loss.

Figure 6 shows the bunch before shaking and Figure 7 shows the bunch after five shakes. It is clear from Figure 6 that the tip of the bunch is moving a lot before it is shaken. After shaking it five times, the bunch distribution has changed and the tip no longer shakes and it has become more rounded than before shaking. Figure 8 superimposes the before and after shaking snapshots which clearly shows the shape change. Closer examination of the bunch current shows that there is no beam loss (to the accuracy of the measurement) during the experiment.

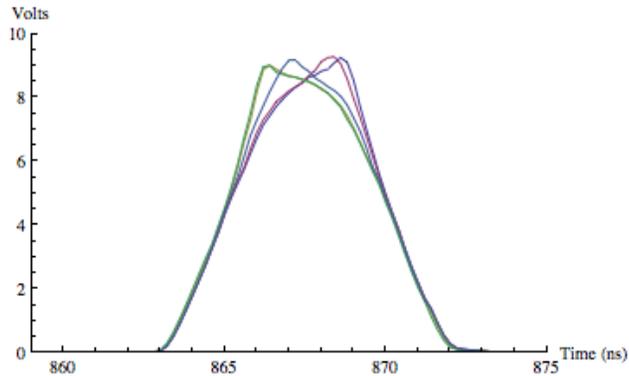

Figure 6: Before any shaking, the bunch tip is moving.

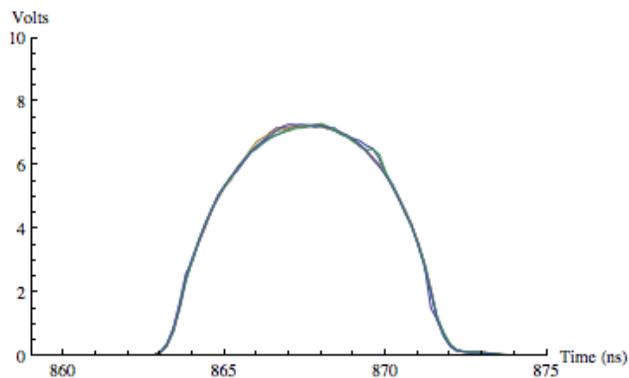

Figure 7: The result after five shakes. Compared to before shaking, the beam has clearly stopped moving and has a more rotund shape.

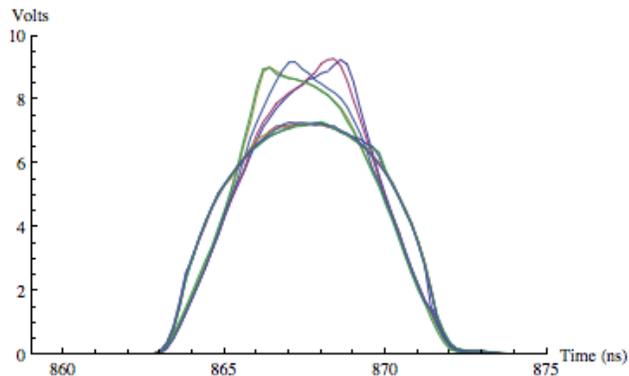

Figure 8: All the traces which are collected before and after shaking are plotted together here. This clearly shows the shape change after shaking.

## CONCLUSIONS

According to predictions of Refs. [2,3], the flattening of the bunch distribution at low amplitudes should make the bunch more stable against LLD. An experiment has been devised to flatten the distribution by modulating the RF phase at the low-amplitude synchrotron frequency for a few degrees of amplitude. These beam studies show that stabilisation really happens. After several consecutive shakings, the dancing disappears and the resulting bunch profile becomes smoother at the top. Although not shown in this report, sometimes a little divot forms at the centre of the distribution. These experiments confirm that resonant RF shaking flattens the bunch distribution at low amplitudes, and the dancing stops.